\documentclass[letterpaper]{article}
\usepackage{aaai19}
\usepackage{times}
\usepackage{helvet}
\usepackage{courier}
\usepackage[hyphens]{url} 
\usepackage{graphicx}
\def\UrlFont{\rm}
\usepackage{graphicx}
\usepackage{makecell}
\usepackage{amssymb}
\usepackage{yhmath}
\usepackage{balance}
\title{Visual Attention Model for Cross-sectional Stock Return Prediction \\and End-to-End Multimodal Market Representation Learning}
\author{
  Ran Zhao \\
  Carnegie Mellon University\\
  \texttt{rzhao1@cs.cmu.edu}
  \\\And
  Yuntian Deng\\
  Harvard University\\
  \texttt{dengyuntian@seas.harvard.edu}
  \\\And
  Mark Dredze\\
  Johns Hopkins University\\
  \texttt{mdredze@cs.jhu.edu}
  \\\AND
  Arun Verma\\
  Bloomberg\\
  \texttt{averma3@bloomberg.net}
  \\\And
  David Rosenberg\\
  Bloomberg\\
  \texttt{drosenberg44@bloomberg.net}
  \\\And
  Amanda Stent\\
  Bloomberg\\
  \texttt{astent@bloomberg.net}
}%
\begin{document}
\maketitle
\begin{abstract} 
Technical and fundamental analysis are traditional tools used to analyze stocks; however, the finance literature has shown that the price movement of each individual stock is highly correlated with that of other stocks, especially those within the same sector. In this paper we propose a general-purpose market representation that incorporates fundamental and technical indicators and relationships between individual stocks. We treat the daily stock market as a `market image' where rows (grouped by market sector) represent individual stocks and columns represent indicators. We apply a convolutional neural network over this market image to build market features in a hierarchical way. We use a recurrent neural network, with an attention mechanism over the market feature maps, to model temporal dynamics in the market. Our model outperforms strong baselines in both short-term and long-term stock return prediction tasks. We also show another use for our market image: to construct concise and dense market embeddings suitable for downstream prediction tasks.

\end{abstract}

\section{Introduction}
\label{sec:intro}
In recent years there have been multiple proposals for methods to adopt machine learning techniques in quantitative finance research.  Modeling stock price movement is very challenging since stock prices are affected by many external factors such as political events, market liquidity and economic strength. However, the rapidly growing volume of market data allows researchers to upgrade trading algorithms from simple factor-based linear regression to complex machine learning models such as reinforcement learning \cite{lee2001stock}, k-nearest neighbors \cite{alkhatib2013stock}, Gaussian processes \cite{mojaddady2011stock} and many deep learning approaches, e.g. \cite{kwon2005stock,rather2015recurrent,singh2017stock}. 

A variety of financial theories for market pricing have been proposed, which can  serve as the theoretical foundation for designing tailored machine learning models. First, the efficient market  hypothesis \cite{malkiel1970efficient} states that all available information is reflected in market prices. Fluctuations in stock prices are a result of newly released information.  Therefore, through analyzing individual stock price movements, a machine learning-based model should be able to decode the embedded market information.  

Second, value investing theory \cite{2008security} suggests to buy stocks below their intrinsic value to limit downside risk. The intrinsic value of a company is calculated by fundamental indicators which are revealed in quarterly and annual financial reports. A machine learning-based model should therefore be capable of discovering the relationships between different types of fundamental indicator and the intrinsic value of a company. 

Third, the methodology of technical analysis introduced in \cite{murphy1999technical} includes  well-known context-dependent leading indicators of price movement such as relative strength index (RSI) and moving average convergence/divergence (MACD). A machine learning-based model should be able to estimate the predictive power of traditional technical indicators in different market situations. 

Fourth, the stock market has a well-defined structure. In the macro, people have invented different financial indexes for major markets such as the NASDAQ-100 and Dow Jones Industrial; these are composite variables that may indicate market dynamics. In the micro, the stock market is usually divided into 10 major sectors and tens of subsectors for key areas of the economy. 
Stocks in the same sector have a shared line of business and are expected to perform similarly in the long run \cite{murphy2011intermarket}.  Traditional ways of dealing with market information are to include hand-crafted  microeconomic indicators in predictive models, or to construct covariance matrixes of returns among groups of stocks. However, those hand-crafted features can become gradually lagged and unable to dynamically adjust to market changes. Therefore, a machine learning-based model should leverage information from the whole market as well as the sector of each included company. 

Inspired by these financial theories, we implement an end-to-end market-aware system that is capable of capturing market dynamics from multimodal information (fundamental indicators~\cite{2008security}, technical indicators~\cite{murphy1999technical}, and market structure) for stock return prediction\footnote{Stock return is appreciation in price (plus any dividends) divided by the original price of the stock.}. First, we construct a  `market image' as in Figure \ref{fig:stock_image}, in which each row represents one stock and each column represents an indicator from the three major categories shown in Table \ref{tab:features}. Stocks are grouped in a fixed order by their sector and subsector (industry). Then we apply state-of-the-art deep learning models from computer vision and natural language processing on top of the market image.  Specifically, our contributions in this work are to: (1) leverage the power of attention-based convolutional neural networks to model spatial relationships between stocks in the market dimension, and of recurrent neural networks for time series forecasting of stock returns in the temporal dimension, and (2) use a convolutional encoder-decoder architecture to reconstruct the market image for learning a generic and compact market representation.

In the following sections, we present our market image, then our models for market-aware stock prediction, and finally our method for computing generic and compact market representations. We present empirical results showing that our model for market-aware stock prediction beats strong baselines and that our market representation beats PCA.

\begin{table}[t]
\centering
\small
    \begin{tabular}{|l|l|c|}
    \hline
    Indicator Set & Time Scale & Indicators  \\
    \hline
    Price-Volume & Daily & \makecell{Close-to-Open ratio, \\High-to-Open ratio,\\Low-to-Open ratio,\\Close-to-High ratio,\\Close-to-Low ratio,\\High-to-Low ratio}  \\
    \hline
    \makecell{Historical\\ Return}& Daily &\makecell{last \{1,2,3,4,5\}-day return,\\last \{5,10,15,20,25,30\}-day\\ cumulative return}  \\
    \hline
    \makecell{Technical\\ Indicators} & Daily & \makecell{BOLL,DMI,RSI,\\MACD,ROC,MOMENTUM}  \\
    \hline
    \makecell{Fundamental \\Indicators} & Quartly & \makecell{EPS,CUR\_RATIO,\\ TOT\_DEBT\_TO\_TOT\_EQY,\\FNCL\_LVGR,\\RETURN\_TOT\_EQY,\\PE\_RATIO,\\SHORT\_INT\_RATIO} \\
    \hline
    \end{tabular}
    \caption {Indicators used in our 'market image'}
    \label{tab:features} 
\end{table}

\section{The Market Image}
\label{sec:models}

\begin{figure}[t]
    \centering
    \includegraphics[width=0.47\textwidth]{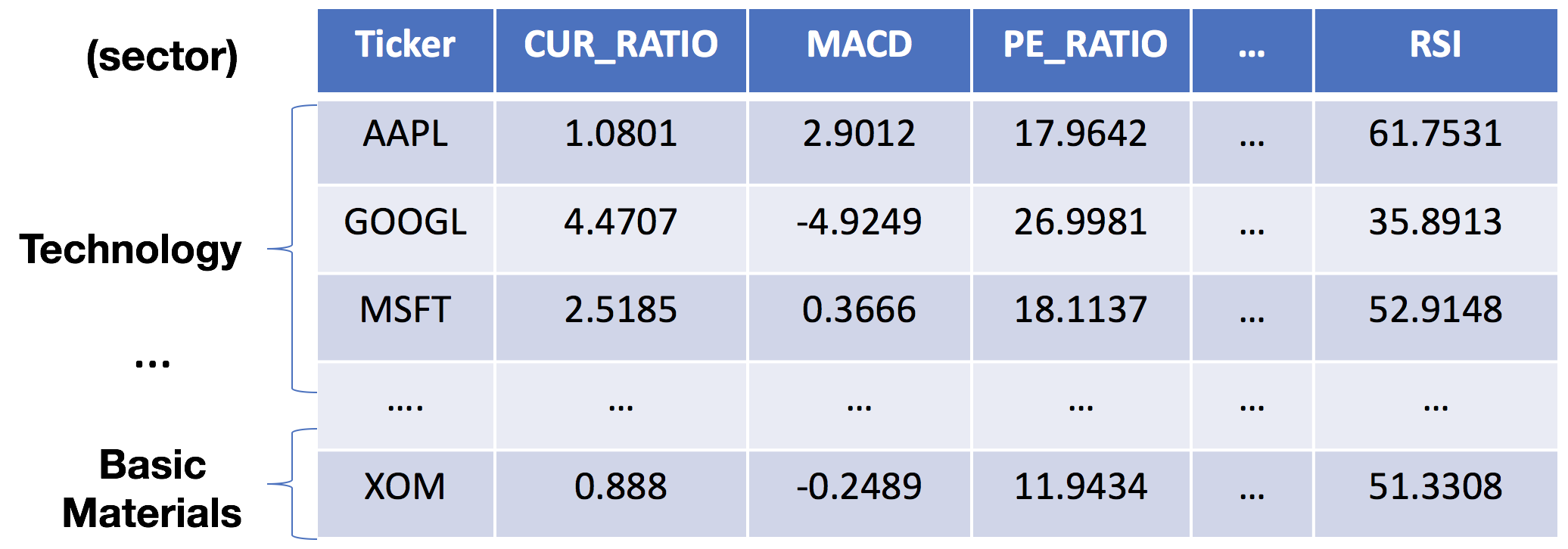}
    \caption{1-day market image snapshot}
    \label{fig:stock_image}
\end{figure}

We represent the daily market as an image $M$, a $m\times n$ matrix where $m$ is the number of unique stocks and $n$ is the number of extracted traditional trading indicators. In our experiments, we used the indicators from Table~\ref{tab:features}. A sample market image is depicted in Figure~\ref{fig:stock_image}. The market image serves as a snapshot of  market dynamics. These market images can be stacked to form a market cube as shown in Figure~\ref{fig:cube}, thus incorporating a temporal dimension into the representation. 

For our experiments later in this paper, we collected the 40 indicators from Table~\ref{tab:features} for each of the S\&P 500 index constituents on a daily basis from January 1999  to Dec 2016, and used this to construct daily market images. The size of the daily image is 500 (stocks) X 40 (multimodal indicators), denoted as  $M_d=\{m_{i=1,...500},n_{j=1,...40}\in \mathbb{R}\}_d$. In each market image, stocks are grouped first by the ten sectors in the Global Industry Classification Standard (GICS), and within each sector, by the GICS subsectors. We normalize the values for each indicator into a 0-1 scale by applying a min-max scalar using the min and max values of that indicator in the training data (see equation (\ref{eq:norm_x})). 
\begin{equation}
  \{M_{i,j'}\}_d=\frac{\{M_{i,j}\}_d-min(\{M_j\}_{\{d\}})}{(max(\{M_j\}_{\{d\}})-min(\{M_i\}_{\{d\}})}
  \label{eq:norm_x}
\end{equation}
Some fundamental indicators are updated quarterly; to fill these blank cells in our market images, we applied a backward fill policy to the affected columns. 

\section{Market-Aware Stock Return Prediction}
\begin{figure}[ht!]
\centering
    \includegraphics[scale=0.6]{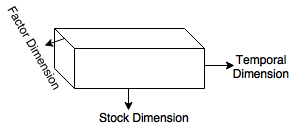}
    \caption{Market Cube}
    \label{fig:cube}
\end{figure}

Let us assume that we want to predict the return of $stock_m$ at day $d$ based on information from the previous $t$ days. 
%The return is measured by the price change of a stock from the close of trading day $d-t$ to $d$.  
This means that we have to learn a market representation with respect to $stock_m$ given the previous $t$ market images as the market context. First we describe our Market Attention model (MA; right side of Figure~\ref{fig:cube}), which builds market-aware representations for individual stocks. Second, we describe how we add temporal modeling to this model to get our Market-Aware Recurrent Neural Network model (MA-RNN; left side of Figure~\ref{fig:cube}). Third, we present empirical results demonstrating that these models outperform strong baselines for stock return prediction.

\subsection{Market Attention Model}

We rotate and stack $t$ market images to construct a 3-D market cube $E$ $\in\mathbb{R}^{t{\times}m{\times}n}$. Rows ($t$) index the temporal dimension, columns ($m$) index stocks, and channels ($n$) index indicators, as shown in Figure~\ref{fig:cube}. Let $x_t^n\in\mathbb{R}^m$ refer to the $m$-dimensional vector indexed by $t$ in the temporal dimension and $n$ in the factor dimension of the market cube $E$ and $y_t^m\in\mathbb{R}^n$ refer to the $n$-dimensional vector indexed by $t$ in the temporal dimension and $m$ in the stock dimension. 

Separately, we initialize stock embeddings $S=\{s^1,s^2,...s^m\}$ to non-zero vectors, where $s^m\in\mathbb{R}^{|v|\times1}$ indexes the $m$-th column's stock embedding.

\begin{figure*}[ht!]
    \centering
    \includegraphics[scale=0.6]{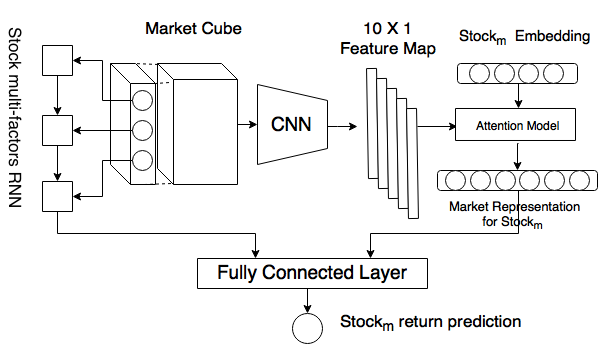}
    \caption{Architecture of Market-Attention Recurrent Neural Network Model (MA-RNN)}
    \label{fig:model1}
\end{figure*}

Then, we use a convolutional neural network (CNN) to generate multiple feature maps of the market cube through multiple convolution operations (right side of Figure \ref{fig:model1}).  Each convolution operation involves a filter $w\in\mathbb{R}^{1\times m}$, which is applied to a window of one day to produce a new feature $c^{j}$ by:
\begin{equation}
    c^{j}=f(\sum^T_{t}\sum^N_{n=1}w^n_{j}\cdot x^{n}_{t}+b), b\in \mathbb{R}
    \label{eq:conv}
\end{equation}
$j$ denotes the j-th kernel; in our experiments, we use 192 different kernels. $f$ is a ReLU active function for introducing non-linearities into the model. So we have a 1-D convolution filter that slides its window vertically with stride=1 along the first dimension of the market cube to produce a feature map column vector $c^j=<c^j_1,c^j_2,...,c^j_{t}>^\intercal$. 

Given a target stock embedding $s^m$, the attention model will return a weighted arithmetic mean of the \{${c^j}$\}, where the weights are chosen according the relevance of each ${c^j}$ to the stock embedding $s^m$. 
We use the additive attention mechanism explained in \cite{attention}. In equation \ref{eq:step1}, $w_{sz}$ and $w_{\bar{c}z}$ are learned attention parameters. 
\begin{equation}
  z_j=\tanh(w_{sz}\cdot s^m+w_{cz}\cdot{c^j})
  \label{eq:step1}
\end{equation}
We compute attention weights using a softmax function
\begin{equation}
  a_j=\frac{\exp(v_j\cdot z_j)}{\sum_i\exp(v_i\cdot z_i)}
  \label{eq:softmax}
\end{equation}
\noindent The conditioned market embedding $p^m$ is calculated by 
\begin{equation}
  p^m=\sum_j a_j{c^j}\\
\end{equation}
Intuitively, each filter serves to summarize correlations between different stocks across multiple indicators.  Each kernel is in charge of finding different types of patterns among the raw indicators.  The attention mechanism of the CNN is responsible for selecting the patterns on which to focus for a particular target stock. The conditioned market embedding summarizes the information contained in the market cube $E$ that is pertinent to the target stock.

\subsection{Market-Aware RNN}

In parallel, we deploy a long-short-term memory recurrent neural network (LSTM) to model the temporal dependencies in a sequence of multidimensional features $y_i^m$of a specific $stock_m$.  Recurrent neural networks \cite{hochreiter1997long,mikolov2010recurrent} are widely applied in natural language processing applications to capture long-range dependencies in time series data \cite{sutskever2014sequence,cho2014properties,dyer2015transition,yang2017neural}. The attention mechanism \cite{attention,luong2015effective} has become an indispensable component in time series modeling with recurrent neural networks; it provides an evolving view of the input sequence as the output is being generated. 
%In our stock market modeling applications, we want to be able to summarize the `trend', while being able to pick up important time points of the past history. Therefore we decided to use recurrent neural networks to model temporal aspects of the stock market. 

The sequence of multidimensional features for $stock_m$ ($y_1^m,y_2^m,...y_t^m$) are sequentially encoded using a LSTM cell of size 25. The mechanism of the LSTM is defined as:

\begin{align*}
  \left[
    \begin{matrix}
      i_t  \\
      f_t  \\
      o_t  \\
      j_t  \\
    \end{matrix}
\right] &= 
  \left[
    \begin{matrix}
      \sigma  \\
      \sigma  \\
      \sigma  \\
      tanh  \\
    \end{matrix}
\right] W [h_{t-1}, x_t]  \\
    c_t &= f_t \odot c_{t-1} + i_t \odot j_t\\
    h_t &= o_t \odot tanh(c_t)
\end{align*}
We treat the last hidden LSTM output, $q^m$, as the representation of the target $stock_m$'s performance in the past $t$ days. 

Finally, we feed both our learned dense market performance embedding $p^m$ and stock performance embedding $q^m$ to a feedforward neural network. They are non-linearly transformed separately in the first layer $\phi$ and concatenated together to predict the target stock return:

\begin{equation}
    f(x)=g(\sum_i W[\phi_1(p^m),\phi_2 (q^m)]+b)
\end{equation}

\subsection{Evaluation}

We conducted an evaluation of our Market-Attention RNN model (MA-RNN). 
For labels, we built stock return matrices for each market image. We used 1-day and 5-day returns for short-term predictions and 15-day and 30-day returns for long-term predictions, denoted as $R_d=\{r_{i=1,...500},_{j=1,...10}\}_d$.
In order to reduce the effect of volatility on returns, we divide the individual daily return by its recent past standard deviation of return ({\it cf} the Sharpe ratio). The moving window size to calculate the standard deviation is 10, see equation (\ref{eq:norm_y}). 
\begin{equation}
\{r_{i,j'}\}_d=\frac{r_{i,j}}{\sigma(r_{j,{\{d-10:d-1\}}})}
\label{eq:norm_y}
\end{equation}

We divided our input market images into training, validation and backtest sets by time, as shown in Table \ref{tab:stat}. 

\begin{table}[ht]
\centering
    \begin{tabular}{l|l|l|l}
    \hline
         &Training & Validation & Backtest \\
         \hline
         Period & 1999-2012 &2012-2015 &2015-2016 \\
         \#Trading Days & 3265 & 754 & 504 \\
         \hline
    \end{tabular}
    \caption{Data Split}
    \label{tab:stat}
\end{table}

We trained our MA-RNN model with the following hyperparameters: a convolution stride size of 1; a  dimensionality of 100 for the trainable stock embeddings; a dimensionality of 32 for the attention vector of the convolutional neural work; a dimensionality of 40 for the final market representation vector; a cell size of 32 for the LSTM; hidden layers of dimensionality 100 and 50 respectively for our fully connected layers; ReLu non-linearity; and a time window $t$ of 10. All the initialized weights were sampled from a uniform distribution [-0.1,0.1]. The mini-batch size was 10. The models were trained end-to-end using the Adam optimizer \cite{kingma2014adam} with a learning rate of 0.001 and gradient clipping at 5. 

For benchmarking our MA-RNN model, we chose several standard machine learning models. We report MSE of the \% return prediction as our metric.

We conducted two experiments. First, we compared the performance of models {\it with} and {\it without} market information. Linear regression (LR), a feedforward neural network\footnote{We used two hidden layers of size 50 and sigmoid non-linearity.} (FFNN), a long-short term memory recurrent neural network (LSTM-RNN) that uses only individual stocks' price histories\footnote{We used a LSTM cell size of 25.} and support vector regression\footnote{We used a linear kernel function with penalty parameter c=0.3.} (SVR) \cite{drucker1997support} serve as our market info-free comparison models. Our Market-Attention model (MA) relies solely on the learned market representation, $p_m$ (with reference to Figure~\ref{fig:model1}, it uses only the CNN with attention, and ignores the output of the LSTM).  We found that {\bf market awareness can be successfully modeled to improve stock return prediction}. As shown in Table \ref{tab:market}, at every time interval (n = 1 day, 5 days, 15 days and 30 days) the Market-Attention (MA) model has lower MSE than the other models, which have no information about the market as a whole\footnote{Obviously, the further away from the current day, the higher the error is expected to be.}. 

\begin{table}[ht]
\centering
    \begin{tabular}{|c|l|l|l|l|}
    \hline
    Model              & n=1 & n=5 & n=15 & n=30  \\
    \hline
    LR & 3.711 &6.750   & 12.381 & 18.429   \\
   SVR  & 2.411 & 4.917 & 8.149 & 11.930   \\
    FFNN           & 1.727 & 3.952 & 6.967 & 9.088   \\
    LSTM-RNN     & 1.426 & 2.896 & 5.854 & 7.923   \\
    MA     & 0.91 & 1.63 & 4.383 & 5.114   \\
    \hline
    \end{tabular}
    \caption{Mean Squared Error of \% Return Prediction}
    \label{tab:market}
\end{table}

Second, we compared the MA model with the full MA-RNN model to show the value of explicitly modeling temporal dependencies. We found that {\bf temporal awareness can be successfully used in a market-aware model for improved stock return prediction}. As shown in Table \ref{tab:memory}, our MA-RNN model has lower MSE than our baseline MA model. 
\begin{table}[ht]
\centering
    \begin{tabular}{|c|l|l|l|l|}
    \hline
    Model              & n=1 & n=5 & n=15 & n=30  \\
    \hline
     MA     & 0.91 & 1.63 & 4.383 & 5.114   \\
     MA-RNN  & 0.790 & 1.210 & 3.732 & 4.523   \\
    \hline
    \end{tabular}
    \caption{Mean Squared Error of \% Return Prediction}
    \label{tab:memory}
\end{table}

\section{Generic Market Representation: MarketSegNet}
\label{sec:exp}

Based on our finding from the previous section that market awareness leads to improved stock prediction accuracy, we propose a novel method to learn a generic market representation (MarketSegNet) in a end-to-end manner. The market representation learning problem is to convert market images (potentially of variable dimensions) to fixed-size dense embeddings for general purpose use.  As a test of the fidelity of this representation, from the generic market embedding it should be possible to reconstruct the input market image pixel wise. 

\begin{figure*}[ht!]
    \centering
    \includegraphics[scale=0.6]{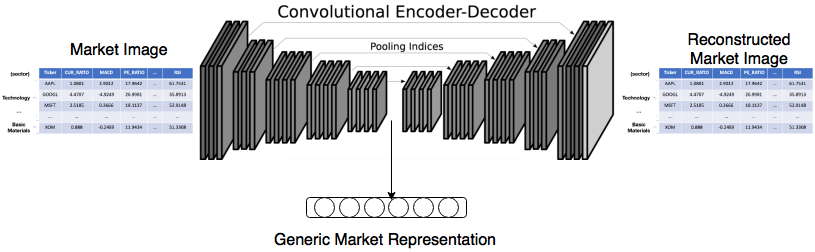}
    \caption{Architecture of MarketSegNet}
    \label{fig:model2}
\end{figure*}

Inspired by \cite{badrinarayanan2017segnet}, we developed a deep fully convolutional autoencoder architecture for pixel-wise regression (Figure~\ref{fig:model2}). The convolutional encoder-decoder model was originally proposed for scene understanding applications, such as semantic segmentation \cite{long2015fully,badrinarayanan2017segnet} and object detection \cite{ren2015faster}. A convolutional encoder builds feature representations in a hierarchical way, and is able to take in images of arbitrary sizes, while a convolutional decoder is able to produce an image of a corresponding size. By using convolutional neural networks, the extracted features exhibit strong robustness to local transformations such as affine transformations and even truncations \cite{zheng2017sift}. In a stock market modeling application, after representing each day's overall stock market as an image, we believe that (1) building features in a hierarchical way can provide a better summary of the market, since stocks exhibit an inherent hierarchical structure, and (2) robustness to local transformations is desirable, since the stock universe is constantly changing, with new companies being added, and other companies removed, while we do not want the overall market representation to be greatly affected by the addition or removal of a single company. 

Since our market image has a different spatial configuration from a normal image, we customize the structure of our end-to-end architecture. The encoder network is composed of traditional convolutional and pooling layers which are used to reduce the resolution of the market image through max-pooling and subsampling operations. Meanwhile, the encoder network stores the max-pooling indices used in the pooling layer, to be applied in the upsampling operation in the corresponding decoder network.
The decoder network upsamples the encoder output using the transferred pool indices to produce sparse feature maps, and uses convolutional layers with a trainable filter bank to densify the feature map so as to recover the original market image. Since companies are grouped in the market image by sector, max-pooling in the encoder network can capture the trends of stocks in the same sector.

To evaluate MarketSegNet, we compare its ability to reconstruct input market images with that of a well-known algorithm for dimensionality reduction, Principal Component Analysis (PCA). PCA uses singular value decomposition to identify variables in the input data that account for the largest amount of variance. We used our training data to train our MarketSegNet model and to fit a PCA model. We then used the MarketSegNet and PCA models to compress and then reconstruct the market images in our test data. We compared the reconstruction error rates of PCA and our MarketSegNet model. Since we varied the sizes of our learned market embeddings from 16 to 128, for each size we created a PCA model with that number of principal components.  

\begin{figure}[ht]
\centering
    \includegraphics[scale=0.3]{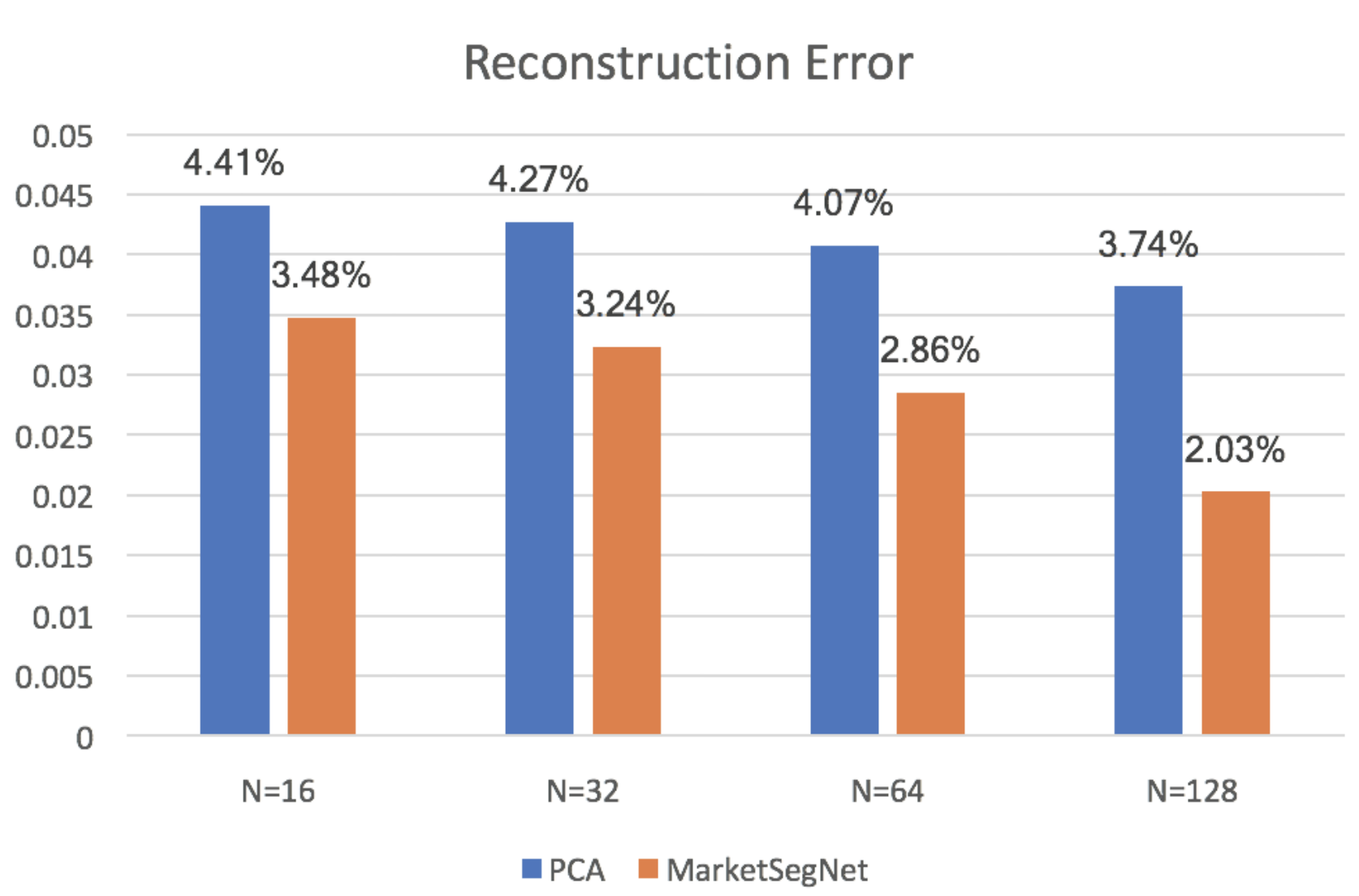}
    \caption{Market Image Reconstruction Error Rates}
    \label{fig:pca}
\end{figure}

Our results are shown in Figure~\ref{fig:pca}. For every size of market embedding, MarketSegNet has lower reconstruction error than PCA.

\section{Conclusions and Future Work}
\label{sec:Conclusion}

In this paper, we present a method for constructing a `market image' for each day in the stock market. We then describe two applications of this market image:
\begin{enumerate}
    \item As input to ML-based models for stock return prediction. We demonstrate (a) that market awareness leads to reduced error vs non-market-aware methods, and (b) that temporal awareness across stacks of market images leads to further reductions in error.
    \item As input to a ML-based method for constructing generic market embeddings. We show that the learned market embeddings are better able to reconstruct the input market image than PCA across a range of dimensionality reductions, indicating that they capture more information about the input market image.
\end{enumerate}

%Our work has contributions to both the finance and AI communities. First, our model quantifies the impact of market indicators (from leading financial theories) on individual stocks.  Second, given the stock market is a leading indicator of economic strength, our model could be applied to monitor and predict market health. 

We should emphasize that our MA model, our MA-RNN model and our MarketSegNet market embeddings do not represent trading strategies. They are agnostic to trading costs, lost opportunity cost while out of the market, and other factors that matter with an active trading strategy. That said, they may provide information that is useful for other AI-driven financial prediction tasks. Other research groups that have used the models described here have reported improved performance in predicting the directionality of stock price moves on earnings day, and in assessing which events will move markets. We leave further exploration of the applications of these models to future work.

% References and End of Paper
% These lines must be placed at the end of your paper
\bibliography{MAIN-F-ZhaoR.82}
\bibliographystyle{aaai}
\end{document}